\documentclass[]{aa}    %

\usepackage{txfonts}
\usepackage{longtable}  
\usepackage{natbib}
\usepackage{graphicx}
\usepackage{deluxetable}

\usepackage{multirow}
\usepackage{placeins}
\usepackage{natbib,amsmath}
\usepackage{graphicx}
\usepackage{hyperref}

\usepackage{float}
\usepackage{caption}
\usepackage{subcaption}
\usepackage{xspace}

\providecommand{\e}[1]{\ensuremath{\times 10^{#1}}}

\usepackage{placeins}
\usepackage{paralist}
\usepackage{graphics,graphicx}

\usepackage{xspace}
\usepackage{graphicx}
\usepackage{lineno}
\usepackage{ulem}

\begin{document}

\newcommand{\ponestring}{13.20362567(12)}
\newcommand{\ptwostring}{17.003479(8)}
\newcommand{\pthreestring}{21.052996(7)}

\titlerunning{TIC\,308396022: A $\delta$ Sct -- $\gamma$ Dor hybrid}
\title{TIC\,308396022: a $\delta$\,Scuti--$\gamma$\,Doradus hybrid with large-amplitude radial fundamental mode and regular g-mode period spacing }

\author{
Tao-Zhi Yang \inst{1} \and Zhao-Yu Zuo \inst{1} \and Gang Li \inst{2} \and Timothy R Bedding \inst{3,4} \and Simon J Murphy \inst{3,4} \and Meridith Joyce \inst{5}
}

\institute{
Ministry of Education Key Laboratory for Nonequilibrium Synthesis
and Modulation of Condensed Matter, School of Physics, Xi'an Jiaotong University, 710049 Xi'an, P.R. China; \\e-mail: zuozyu@xjtu.edu.cn (ZYZ);\\ 
\and
IRAP, Universit\'{e} de Toulouse, CNRS, CNES, UPS, Toulouse, France;\\
\and
Sydney Institute for Astronomy (SIfA), School of Physics, University of Sydney, Camperdown NSW 2006, Australia\\
\and
Stellar Astrophysics Centre, Department of Physics and Astronomy, Aarhus University, Ny Munkegade 120, DK-8000 Aarhus C, Denmark\\
\and
Space Telescope Science Institute in Baltimore, MD, USA;\\
}


\abstract
{
We analyse the pulsating behaviour of TIC\,308396022 observed by the TESS mission. The star is a high-amplitude $\delta$\,Sct star (HADS) that shows a very rich amplitude spectrum using the 3-yr light curve. Among these frequencies, the strongest peak of $f_{1}= \ponestring\,\rm{d^{-1}}$ is identified as the radial fundamental mode, and we also find the first and second overtones ($f_2$ and $f_3$). In the low frequency range (< 2.5 $\rm{d^{-1}}$), 22 peaks are identified to be gravity modes, which show a regular period spacing of about 2460\,s and have the angular degree $l = 1$. The period spacing pattern does not show a significant downward trend, suggesting the star rotates slowly. We note that this is a $\delta$\,Sct--$\gamma$\,Dor hybrid star containing a high-amplitude radial fundamental mode and a regular g-mode period spacing pattern. With O-C analysis, we find the star shows a significant time delay, implying that the star has a companion which is likely to be a white dwarf. The history of possible mass transfer provides a great opportunity to test the current theories of binary evolution, mass transfer, and pulsation. 

}

\keywords{stars: oscillations; stars: variable: delta Scuti; individual: TIC\,308396022}

  \maketitle

\section{Introduction}

The study of stellar oscillations is a powerful tool to probe the internal structures of stars by using stellar periodic oscillations \citep{2010aste.book.....A,2013ARA&A..51..353C,2015pust.book.....C,2021RvMP...93a5001A}. Oscillations can be observed photometrically \citep{2013pss4.book..207H,2021RvMP...93a5001A} and pulsation frequencies can be found by calculating the amplitude spectrum \citep[e.g.][]{Lomb1976,Scargle1982}. After that, many parameters, such as mass, age, rotation, and distance can be measured, which are not straightforward to observe otherwise. 

A class of short-period pulsating stars, $\delta$\,Sct stars, shows great potential for asteroseismic studies. They are intermediate-mass variables with spectral types between A2 and F2, and situated in the lower end of the classical instability strip \citep{Breger2000,2019MNRAS.485.2380M}. Most $\delta$\,Sct stars are multi-periodic pulsators and may show radial and non-radial pulsations \citep{2017ampm.book.....B}. These pulsations are driven by the $\kappa$ mechanism operating in the He-II partial ionization zone \citep{Breger2000,2010aste.book.....A,2014ApJ...796..118A,2020MNRAS.498.4272M}, therefore the pulsations are low-radial-order ($n$) low-angular-degree ($l$) pressure (p) modes with periods between 15 minutes and 8 hours \citep{2011A&A...534A.125U,2014MNRAS.439.2078H}. Their masses are generally between $1.5 \mathrm{M_\odot}$ and $2.5 \mathrm{M_\odot}$, which place them in the transition region, the lower mass stars with a thick outer convective envelope and the massive cases with thin convective shell \citep{2017ampm.book.....B}. Thus, the pulsations in $\delta$\,Sct stars provide an opportunity to make detailed studies for the structure and evolution of stars in this transition region \citep{2018MNRAS.476.3169B}.   

However, the amplitude spectra of $\delta$\,Sct stars are generally very rich and messy, which challenges the mode identification \citep{Goupil_2005, Handler_2009_challenges}. Recently, \cite{Bedding2020Natur} found that some young multi-periodic $\delta$\,Sct stars show regular frequency separations as predicted by the asymptotic relation \citep{Shibihashi1979, 1980ApJS...43..469T}, providing a new possibility for asteroseismology in $\delta$\,Sct stars. In addition, there is a sub-class $\delta$\,Sct stars named high-amplitude $\delta$\,Sct stars (HADS), whose mode identification is relatively clear. The HADS usually pulsate in the radial fundamental mode and overtones, with amplitudes of about $0.1$ magnitude \citep{Petersen_1989, Petersen_1996}. The target in this work, TIC\,308396022, is a HADS.

A- to F-type main-sequence (MS) stars may also pulsate in gravity (g) modes. These are known as $\gamma$\,Dor stars, which appear near the red edge of the $\delta$\,Sct instability strip and show low-frequency light variations \citep{Balona1994,Kaye1999,Dupret2005}. These stars have typical masses from 1.4 to 2.0\,$\mathrm{M_\odot}$ \citep[e.g.][]{Kaye1999, VanReeth2016}, and they usually pulsate in low-degree high-order g modes driven by the convective blocking mechanism \citep{2000ApJ...542L..57G,2004A&A...414L..17D, Dupret2005}, with periods between 0.2 days and 2 days and typical amplitudes below 0.01 mag \citep{LiGang_2020_611}. The g modes sometimes show uniform spacings in period, as predicted by the asymptotic relation. Usually, the uniform period spacings contain the information of the inner chemical composition gradients and the stellar evolutionary status \citep[e.g. ][]{2008MNRAS.386.1487M,2018ApJ...867...47W,2015A&A...580A..27M,2019ApJ...881...86W,2020ApJ...899...38W,2021A&A...648A..91S}. However, for some special case, they also contain other information, such as,  near-core rotation rates \citep{Bouabid2013, Ouazzani2017}, and coupling between g modes and inertial modes \citep{Ouazzani2020, Saio2021}.

The overlapping of the instability strips of $\delta$\,Sct and $\gamma$\,Dor stars allows the existence of hybrid stars that show both p- and g-mode pulsations \citep{1996A&A...313..851B}. The g modes carry the information of the near-core region and the p modes can probe the stellar envelope \citep{2010ApJ...713L.192G, Kurtz_2014_11145123, Saio_2015_9244992}, so the hybrid pulsators of $\delta$\,Sct and $\gamma$\,Dor stars have great potential to contribute to our understanding of  the internal structure of a star \citep[see][]{Kurtz_2014_11145123,Saio_2015_9244992,2016A&A...592A.116S,2017A&A...597A..29S}. Several $\delta$\,Sct--$\gamma$\,Dor hybrid stars have been detected by ground-based observations \citep{2002MNRAS.333..251H,2005AJ....129.2026H,2009MNRAS.398.1339H}, and the large number observed by space missions such as \textit{CoRoT} \citep{2009IAUS..253...71B} and \textit{Kepler} \citep{Borucki2010} suggested that hybrid behavior might be common in A-F stars \citep{2010ApJ...713L.192G,2010arXiv1007.3176H}. Based on a large sample (750 stars) of $\delta$\,Sct and $\gamma$\,Dor candidates discovered in Kepler mission, \citet{2011A&A...534A.125U} found that about 63\% (471 stars) of the sample show $\delta$\,Sct or $\gamma$\,Dor pulsations, and 36\% (171 stars) are hybrid $\delta$\,Sct--$\gamma$\,Dor stars. Recent studies \citep[e.g.][] {2015AJ....149...68B} with a larger sample of $\delta$\,Sct--$\gamma$\,Dor stars suggest that hybrid stars are very common. With the successful launch of TESS mission \citep{2014SPIE.9143E..20R,Ricker2015}, more hybrid stars will be found and hence provide better understandings for the inner structures and oscillation spectra of this kind of variables. 

TIC\,308396022 (TYC 8928-1300-1; $\alpha_{2000}$=$08^{h}$$01^{m}$$02^{s}$.370, $\delta_{2000}$=$-63^{\circ}$$40^{'}$$30^{''}$.317) was first observed in the Sector 1 of the TESS observations and classified to be a pulsating star in \citet{2019MNRAS.490.4040A}. Its basic properties from that study and the TESS Input Catalogue \citep[TIC;][]{2018AJ....156..102S} are listed in Table \ref{tab:basic_parmeters}. To investigate the pulsations of TIC\,308396022 further, we analyse the 2-min cadence photometric data spanning three years from the TESS mission.

\begin{deluxetable}{lcc} 

\tabletypesize{\small} 
\tablewidth{0.5\textwidth} 
\tablenum{1} 
\tablecaption{Basic parameters of TIC\,308396022.} 
\tablehead{ \colhead{Parameters} & \colhead{TIC\,308396022} & \colhead{References} }
\startdata 
   TESS magnitude              &  11.069  &   a  \\
   alternative ID  &  GSC 08928-01300      &   a  \\
                           &  2MASS J08010239-6340302 &  \\
   $T_\mathrm{eff}$            &  6730 $\pm$ 248 K     &  (Gaia), a  \\
                        &  6860 $\pm$ 150 K     &   SED, b  \\
                        &  7371 $\pm$ 150 K     &   SED, c  \\
   log $g$              &  3.81 $\pm$ 0.25 dex  & (Gaia, phot), a  \\
                        &  4.12 $\pm$ 0.25 dex  &  c \\
   log ($L/L_{\odot}$)  &  0.94 $\pm$ 0.01      &  (Gaia), a  \\
   Parallax (mas)       &  1.71  $\pm$ 0.02     &  (Gaia), a \\
   B     &  11.55       &   c \\
   V     &  10.997      &   c\\
   J     &  10.573      &   c \\
   H     &  10.471      &   c \\
   K     &  10.371      &   c \\
   Gaia mag &  11.330   &   a \\
   
   \enddata 
   
   \tablecomments{(a) Gaia \citep{2017MNRAS.471..770M}. (b) \citet{2019MNRAS.490.4040A}. (c) Parameters from the TESS Input Catalogue \citep{2018AJ....156..102S}: https://tasoc.dk/catalog/.}
\label{tab:basic_parmeters}
\end{deluxetable}

\begin{figure}
\begin{center}
  \includegraphics[width=0.5\textwidth,trim=10 10 10 10,clip]{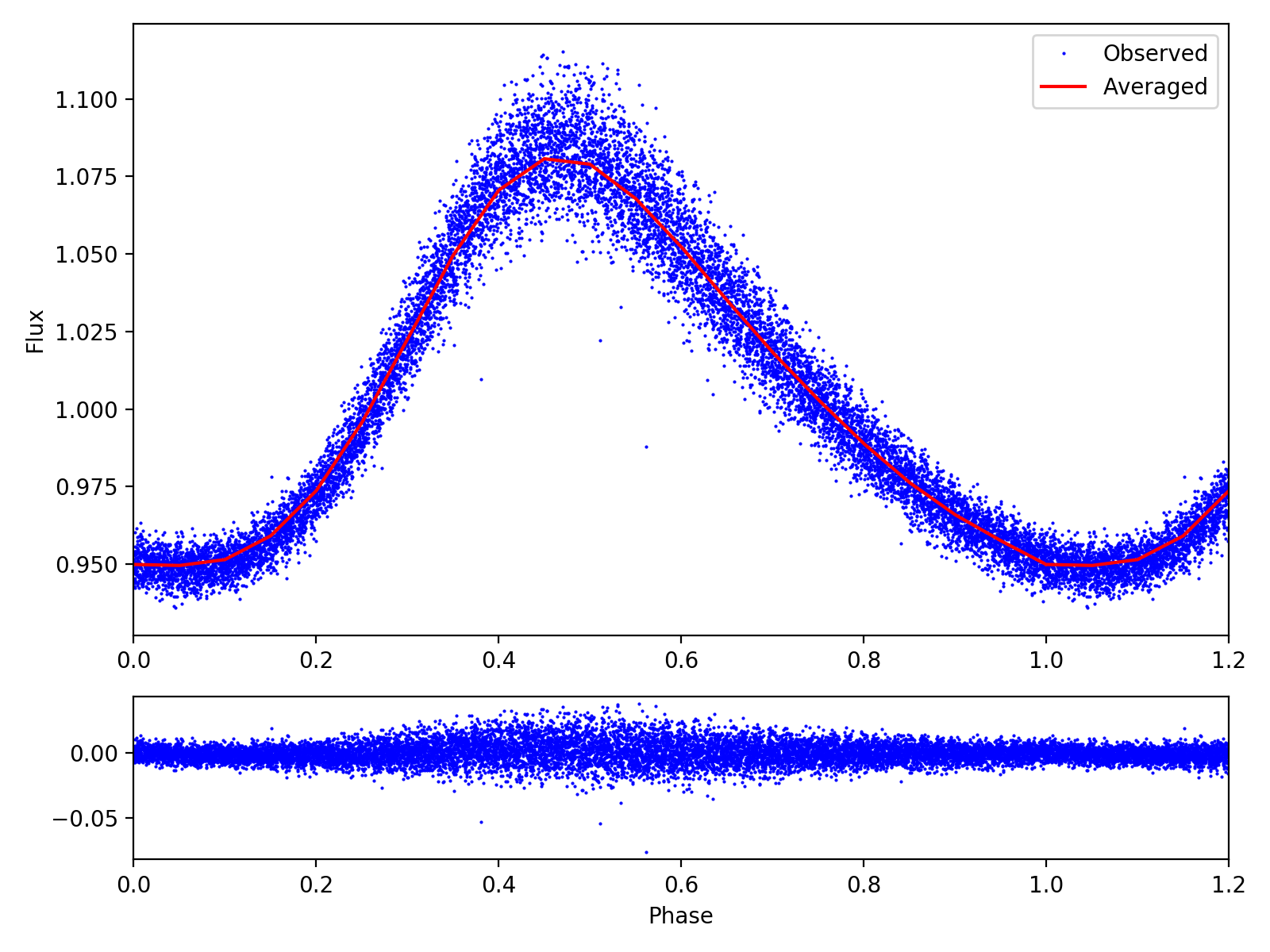}
  \caption{The phase diagram of TIC 308396022 from TESS data, folded by the frequency $f_{1}$ =13.20362567 d$^{-1}$. The obvious light variation of rapidly climbing and slow descent is typical for a HADS star.}
    \label{fig:phase_diagram}
\end{center}
\end{figure}

\section{Observation and Data Reduction}

TIC\,308396022 was observed in Sectors 1, 4, 5, and 7 -- 11 during the first cycle of the TESS mission, and again in Sectors 27, 28, 31, 34, 35, 37, and 38 during Cycle 3. We use all these light curves, which are available from the Mikulski Archive for Space Telescopes (MAST) \footnote{MAST: {http://archive.stsci.edu/}}. We downloaded all the data and used the 2-min cadence Simple Aperture Photometry (SAP) light curves for our research. In each sector, the obvious outliers were removed and the slow trend was corrected with a high pass filter. In total, the rectified light curve includes 236991 data points spanning 1035 days with a duty-cycle of $32\%$. Figure~\ref{fig:phase_diagram} shows the phased light curve, folded by the dominant frequency \textbf{$f_1=\ponestring\,\mathrm{d^{-1}}$}. The peak-to-peak amplitude of the dominant mode is $\sim14.7\%$.

\section{Frequency Analysis and Mode Identification}

To extract the pulsation frequencies, we calculated the Fourier transform of the reduced light curve using the software PERIOD04 \citep{Lenz2005}. The peaks are extracted one-by-one via the method of pre-whitening, in the processes, the light curve is fitted with the following formula:
\begin{equation}
m = m_{0} + \Sigma\mathnormal{A}_{i}\sin(2\pi(\mathnormal{f}_{i}\mathnormal{t} + \phi_{i})), \label{equation1}
\end{equation}
where $m_{0}$ is the zero-point, $A_{i}$, $f_{i}$, and $\phi_{i}$ are the amplitude, frequency, and phase, respectively, of  the \textit{i$^{th}$} peak.

We extracted all the frequencies with $\mathrm{S/N}$ > 4.0 \citep{Breger1993} and $f$ < 80 d$^{-1}$, which covers the typical pulsation frequency range of $\delta$\,Sct stars. The uncertainties of frequencies were calculated following \citet{1999DSSN...13...28M} and \citet{2012IAUS..285...17K}:
\begin{equation}
    \sigma_f=0.44\frac{1}{T}\frac{1}{\mathrm{S/N}},
\end{equation}
where $T=1035\,\mathrm{d}$ is the observation time span, and $\mathrm{S/N}$ is the signal-to-noise ratio of the peak. The noise is the mean level within 2.5 c/d around the peak.
  
\subsection{Pressure modes}

Based on the Gaia EDR3 distance of $591 \pm11$\,pc \citep{2021AJ....161..147B} and the apparent magnitude of $V=11.0$, we calculated the absolute magnitude to be $M_V = 2.14 \pm 0.04$ (neglecting extinction). According to the period--luminosity relation from \citet{2019MNRAS.486.4348Z}, the frequency of the radial fundamental mode is expected to be about $13.86 \pm 0.43$\,d$^{-1}$. The dominant frequency of $f_1=\ponestring\,\mathrm{d^{-1}}$ is therefore consistent with the radial fundamental, especially given the intrinsic scatter in the P--L relation. In fact, if we consider a small extinction ($A_V=0.05$) for calculating the absolute magnitude $M_V$, then the P--L relation predicts the radial fundamental mode to be $13.3\pm0.4$ \,d$^{-1}$, in agreement with the observed mode.

\begin{figure}
    \centering
    \includegraphics[width=0.5\textwidth]{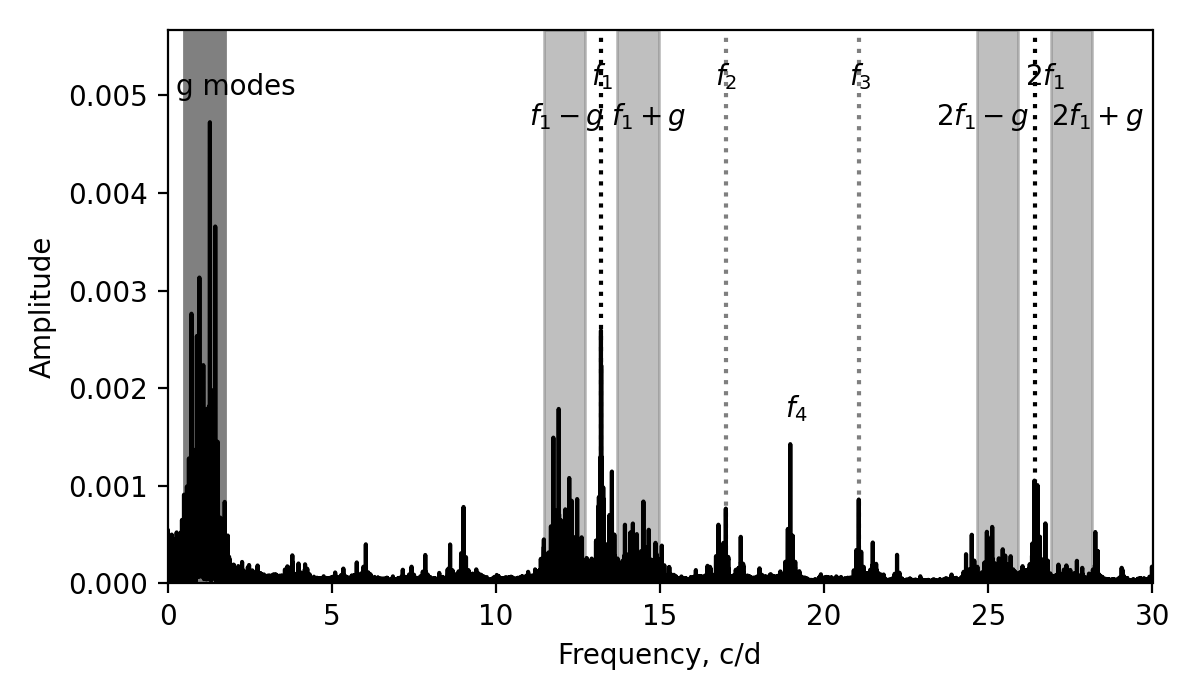}
    \caption{The amplitude spectrum of TIC\,308396022, with the fundamental mode removed. The vertical dark-gray dotted line marks the location of the fundamental mode $f_1$, and the light-gray dotted lines mark the first and the second overtones, $f_2$ and $f_3$. The dark shaded area shows the g-mode region, while the light shaded areas around $f_1$ and 2$f_1$ show the combinations between the fundamental mode and the g modes: $f_1+g$, $f_1-g$, 2$f_1+g$, and 2$f_1-g$.}
    \label{fig:FFT}
\end{figure}

The radial fundamental mode contributes the majority of the light variations. To see more details in the power spectrum, we removed the fundamental mode by the mean phased light curve shown in Fig.~\ref{fig:phase_diagram}, and in Fig.~\ref{fig:FFT} we display the amplitude spectrum of the residuals. We still can see the residual of the strongest fundamental mode with frequency $f_{1}=\ponestring\,\mathrm{d^{-1}}$. As discussed in Sec. 5, these are sidelobes split by the obital frequency \citep{Shibahashi2012,2015MNRAS.450.3999S}. We note that a peak appears at $f_2=\ptwostring\,\mathrm{d^{-1}}$, with a ratio of $f_1/f_2=0.7765$. Therefore, we identify the mode at $f_2$ as the first overtone. The second overtone is also found at $f_3=\pthreestring\,\mathrm{d^{-1}}$ with a ratio of $f_1/f_3=0.6272$ \citep{1979ApJ...227..935S,2021arXiv210708064N}. \textbf{The p-mode frequencies and mode identifications are listed in Table~\ref{tab:p-mode}. When assigning the radial orders, we followed the convention advocated by \citet{2000ASPC..203..529G} that the fundamental mode is $n=1$, the first overtone is $n=2$ and the second overtone is $n=3$.}

Another stronger peak \textbf{$f_{4}$} (=18.96664 d$^{-1}$) has a ratio of \textbf{$f_{1}$/$f_{4}$} = 0.696, which is not equal to any value of the period ratios of the first four radial mode for $\delta$\,Sct stars. Therefore, \textbf{$f_{4}$} must belong to a non-radial mode. Given that $f_{4}$ is very close (within 0.33 \%) to halfway between $f_2$ (the first radial overtone) and $f_3$ (the second radial overtone), it is likely to be an $l=1$.

\begin{deluxetable}{cccc} 
\tabletypesize{\small} 
\tablewidth{0.5\textwidth}
\tablenum{2} 
\tablecaption{The frequencies and identifications of the p modes in TIC 308396022. \label{tab:p-mode}} 
\tablehead{ \colhead{Name} & \colhead{Frequency, $\mathrm{d^{-1}}$} & \colhead{radial order (n)} & \colhead{angular degree (l)}}
\startdata 
   $f_1$ & 13.20362567(12) & 1 & 0\\
    $f_2$ & 17.003479(8)   & 2 & 0\\
    $f_3$ & 21.052996(7)  & 3 & 0\\
    $f_4$ & 18.96664(1)   & - & 1\\
   
   \enddata 
\end{deluxetable}

\begin{deluxetable}{llc} 
\tabletypesize{\small} 
\tablewidth{0.5\textwidth} 
\tablenum{3} 
\tablecaption{The g-mode frequencies and periods. The numbers in the brackets show the uncertainties in the last digits. The radial order is calculated by the period divided by the period spacing $\Delta P$, which is set $\Delta P = 2460\,\mathrm{s}$. \label{tab:g_mode}} 
\tablehead{ \colhead{Frequency, $\mathrm{d^{-1}}$} & \colhead{Period, $\mathrm{d}$} & \colhead{Radial order} }
\startdata 
   2.05778(6) & 0.485961(14) & 17 \\ 
1.94024(8) & 0.515400(20) & 18 \\ 
1.834369(18) & 0.545147(5) & 19 \\ 
1.740762(11) & 0.574461(4) & 20 \\ 
1.657888(19) & 0.603177(7) & 21 \\ 
1.58337(3) & 0.631564(13) & 22 \\ 
1.4524286(23) & 0.6885020(11) & 24 \\ 
1.394985(9) & 0.716853(5) & 25 \\ 
1.340569(14) & 0.745952(8) & 26 \\ 
1.2899600(16) & 0.775218(10) & 27 \\ 
1.243789(25) & 0.803995(16) & 28 \\ 
1.16131(4) & 0.861095(26) & 30 \\ 
1.124017(7) & 0.889666(5) & 31 \\ 
1.088470(5) & 0.918721(4) & 32 \\ 
1.05770(4) & 0.94545(3) & 33 \\ 
1.025381(16) & 0.975247(15) & 34 \\ 
0.969583(5) & 1.031371(5) & 36 \\ 
0.894606(3) & 1.117810(4) & 39 \\ 
0.793373(7) & 1.260441(11) & 44 \\ 
0.77478(5) & 1.29069(8) & 45 \\ 
0.743097(15) & 1.345720(28) & 47 \\ 
0.7280023(29) & 1.373622(6) & 48 \\ 
   \enddata 
\end{deluxetable}

\subsection{Gravity modes}
The modes with frequencies below $\sim2.5\,\mathrm{d}^{-1}$, shown as the dark shaded area in Fig.~\ref{fig:FFT}, are below the typical frequency range of $\delta$\,Sct stars. We also see the combinations between the fundamental mode $f_1$ and these low frequencies, marked by the light shaded areas in Fig.~\ref{fig:FFT}, which proves that they originate from the same star. 

\begin{figure}
\begin{center}
  \includegraphics[width=0.5\textwidth,trim=10 0 35 0,clip]{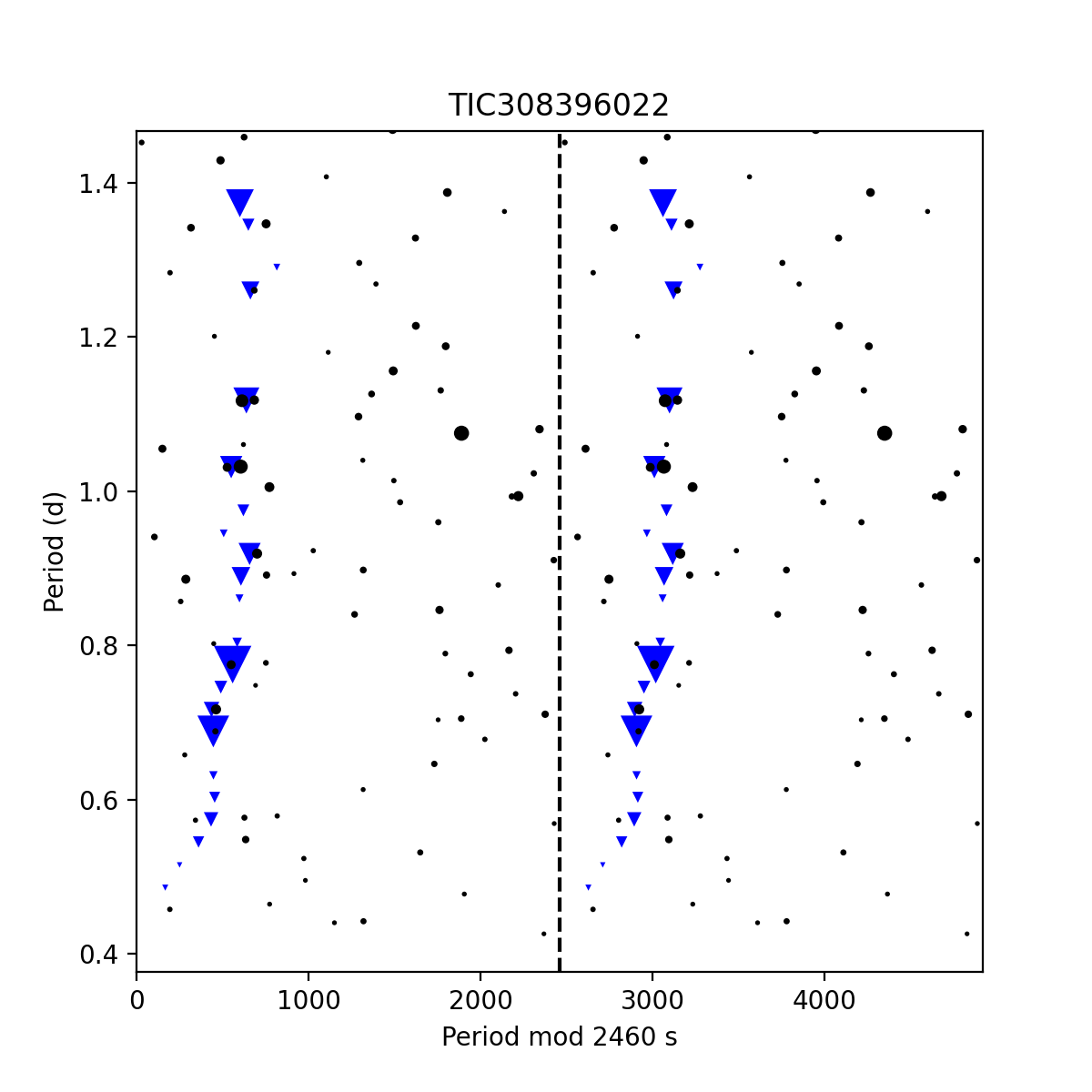}
  \caption{The period $\rm{\acute{e}}$chelle diagram of TIC\,308396022, showing a clear period spacing of about $2460\,\mathrm{s}$. The blue triangles indicate g modes with $l$=1, while the black dots are probably noise peaks.}
    \label{fig:echelle_diagram}
\end{center}
\end{figure}

\begin{figure}
\begin{center}
  \includegraphics[width=0.5\textwidth,trim=0 0 0 0,clip]{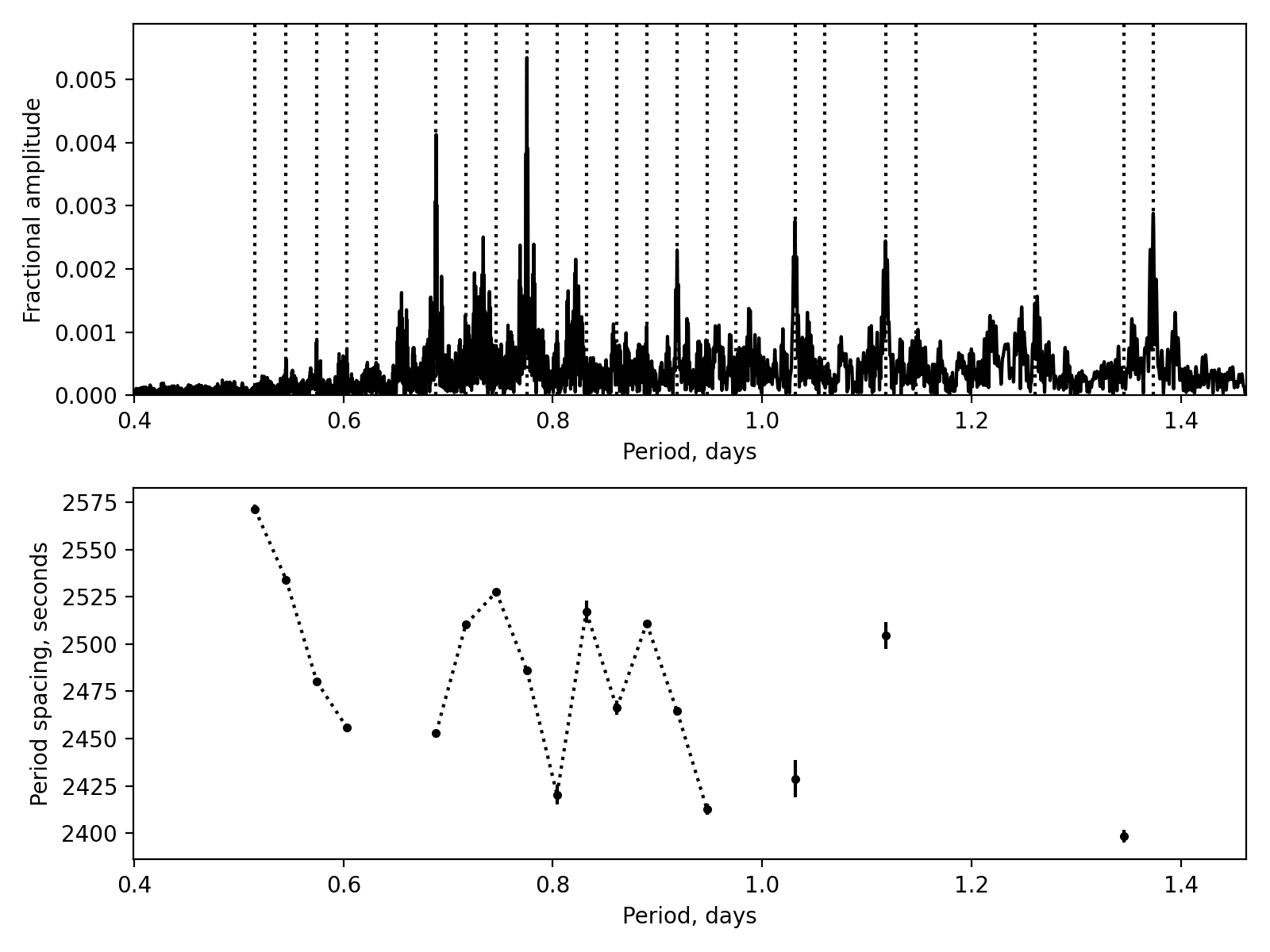}
  \caption{The amplitude spectrum and period spacing patterns of TIC\,308396022.  Upper panel: the amplitude spectrum with x-axis of period. The vertical dashed lines show the locations of the g modes in the period spacing patterns. Lower panel: the period spacing pattern. The dots are the period spacings between the adjacent modes.}
    \label{fig:subplot}
\end{center}
\end{figure}

To identify these modes, we made the period \'{e}chelle diagram for TIC\,308396022, following the method described in \citet{2015EPJWC.10101005B} and \citet{LiGang2019}. The result is shown in Fig.~\ref{fig:echelle_diagram} and Table~\ref{tab:g_mode}. It clearly shows the properties of the uniform period spacing with a period spacing $\Delta P$ of about $2460\,\mathrm{s}$, which is within the typical range for dipole $l=1$ g modes in $\gamma$\,Dor stars \citep{VanReeth2015,VanReeth2016,2019MNRAS.482.1757L}. We show the period spacing pattern in Fig.~\ref{fig:subplot}, and find that the period spacings $\Delta P$ fluctuate around 2500\,s, but do not show a clear downward trend. The obvious fluctuation implies that TIC\,308396022 might be an evolved star and far from the zero-age main-sequence (ZAMS) \citep{2008MNRAS.386.1487M,2018ApJ...867...47W}. Following the method by \cite{VanReeth2016} and \cite{LiGang2019}, we obtain the asymptotic spacing as $\Pi_0 = \sqrt{l(l+1)}\Delta P = 3655\pm13\,\mathrm{s}$, which is smaller than the typical value of the $\gamma$\,Dor stars \citep[about 4000\,s, see][]{LiGang_2020_611}. The small asymptotic spacing value also shows that this star has evolved to the end of the MS. However, the flat period spacing pattern indicates that the near-core rotation is slow, so that the effect of the Coriolis force is negligible \citep{Bouabid2013}. By the method from \cite{VanReeth2016} and \cite{LiGang2019}, we estimate the rotation rate as $0.006\pm0.003\,\mathrm{d^{-1}}$. Note that the slope in the $\Delta P - P$ relation is very small and it is strongly influenced by the scatter rather than the rotation effect, hence the rotation rate we give is imprecise. Slow-rotation is also consistent with the suggestion that the only difference between the HADS and small-amplitude $\delta$\,Sct stars is the rotation rate \citep{Xiong2016}. 

\section{The location in the H-R Diagram}\label{sec:locationHR}

\begin{figure}
\begin{center}
  \includegraphics[width=0.5\textwidth,trim=10 10 10 10,clip]{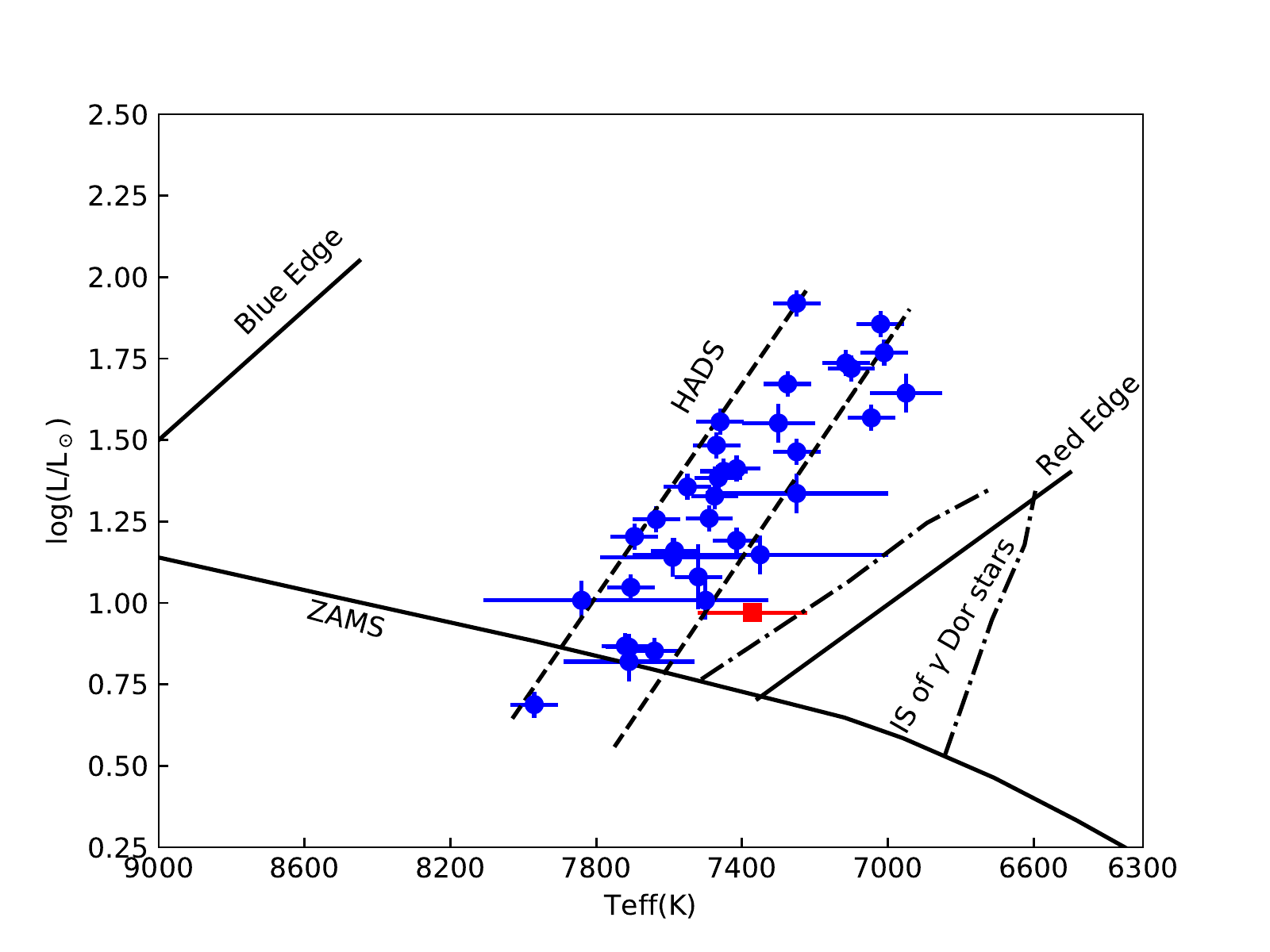}
  \caption{Location of the 34 well-studied HADS and TIC\,308396022 in the H-R Diagram. TIC\,308396022 is shown as the red square. The blue dots are the HADS collected from \cite{McNamara2000,Poretti2005,Poretti2011,Christiansen2007,Balona2012a,Ulusoy2013,Pena2016,Yang2018b,Yang2019,2021MNRAS.504.4039B}. The zero-age main-sequence (ZAMS) and the $\delta$\,Sct instability strip (solid lines) are from \citet{2019MNRAS.485.2380M}. The dashed lines show the region occupied by HADS, as found by \cite{McNamara2000}. The theoretical instability strip (IS) of $\gamma$ Dor stars (dotted-dashed lines) is from \cite{2005A&A...435..927D}.}
    \label{fig:HR_diagram}
\end{center}
\end{figure}

To investigate the evolutionary state of TIC\,308396022, we calculated its luminosity as: L= 9.31 $\pm$ 0.50 $L_{\odot}$ (log $L/L_{\odot}$ = 0.97 $\pm$ 0.02) using the parameters from TIC \citep{2018AJ....156..102S}. Based on the luminosity log $L/L_{\odot}$ = 0.97 $\pm$ 0.02 and $T_\mathrm{eff}$ = $7371 \pm150$\,K from the TIC, we plot the location of TIC\,308396022 in the H-R Diagram shown in Fig.~\ref{fig:HR_diagram}, as well as additional 34 well-studied HADS stars collected from literature \citep{McNamara2000,Poretti2005,Poretti2011,Christiansen2007,Balona2012a,Ulusoy2013,Pena2016,Yang2018b,Yang2019}. From this figure, TIC\,308396022 lies between the $\delta$\,Sct and the $\gamma$\,Dor instability strip on the MS. We note that both the effective temperatures derived from $Gaia$ and \cite{2019MNRAS.490.4040A} are lower than that from TIC, which makes TIC 308396022 locate far apart from the HADS region and $\delta$\,Sct instability strip.

\begin{figure}
    \centering
    \includegraphics[width=1\linewidth]{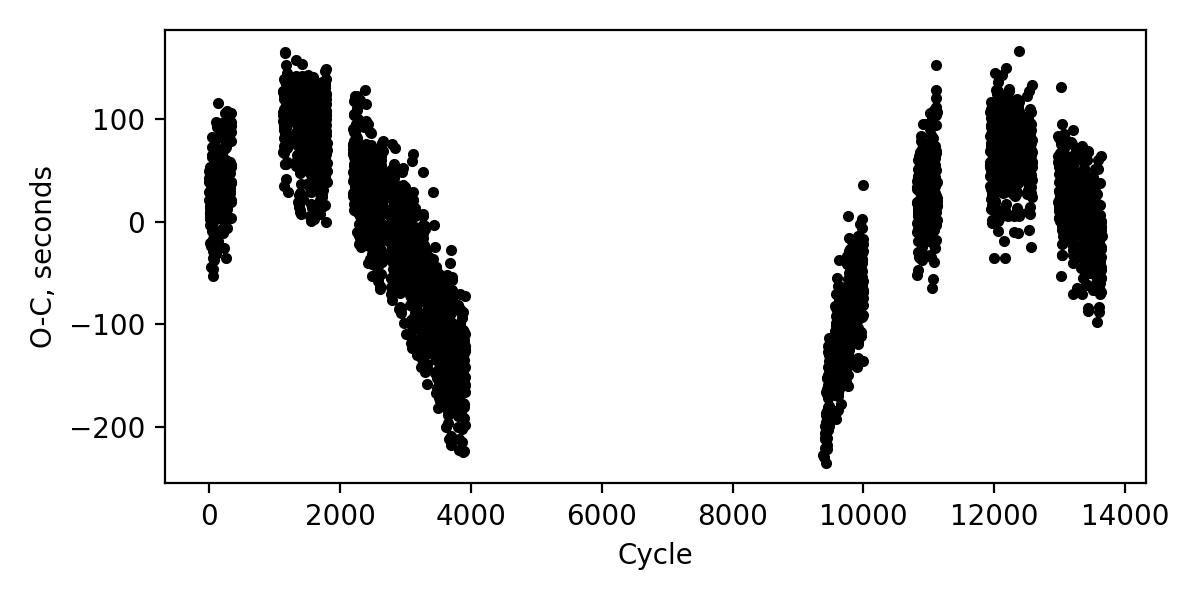}
    \caption{The $O-C$ diagram of TIC\,308396022, based on the dominant pulsation mode (\textbf{$f_1$}).}
    \label{fig:O-C}
\end{figure}

\begin{figure}
    \centering
    \includegraphics[width=1\linewidth]{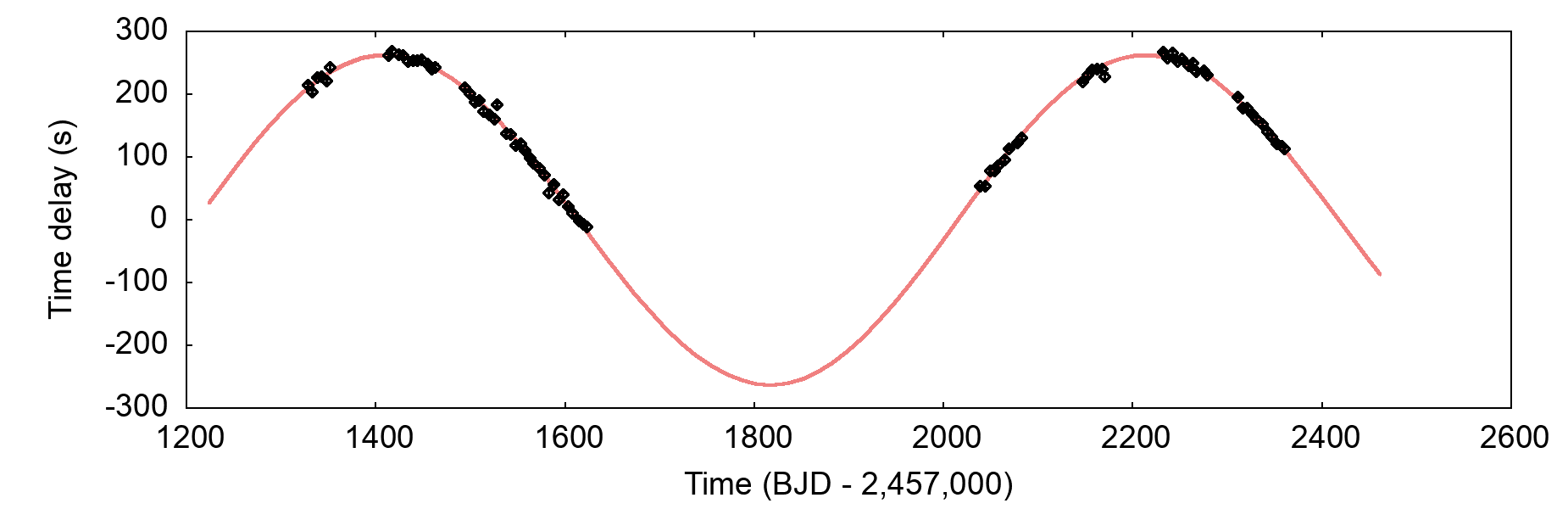}
    \caption{Time delay measurements (black symbols) for TIC\,308396022, using 10-d light curve segments. Overlapping red curves show 25 random samples from the Markov chain that are indistinguishable of this resolution.}
    \label{fig:time_delay}
\end{figure}

\begin{figure}
    \centering
    \includegraphics[width=0.95\linewidth]{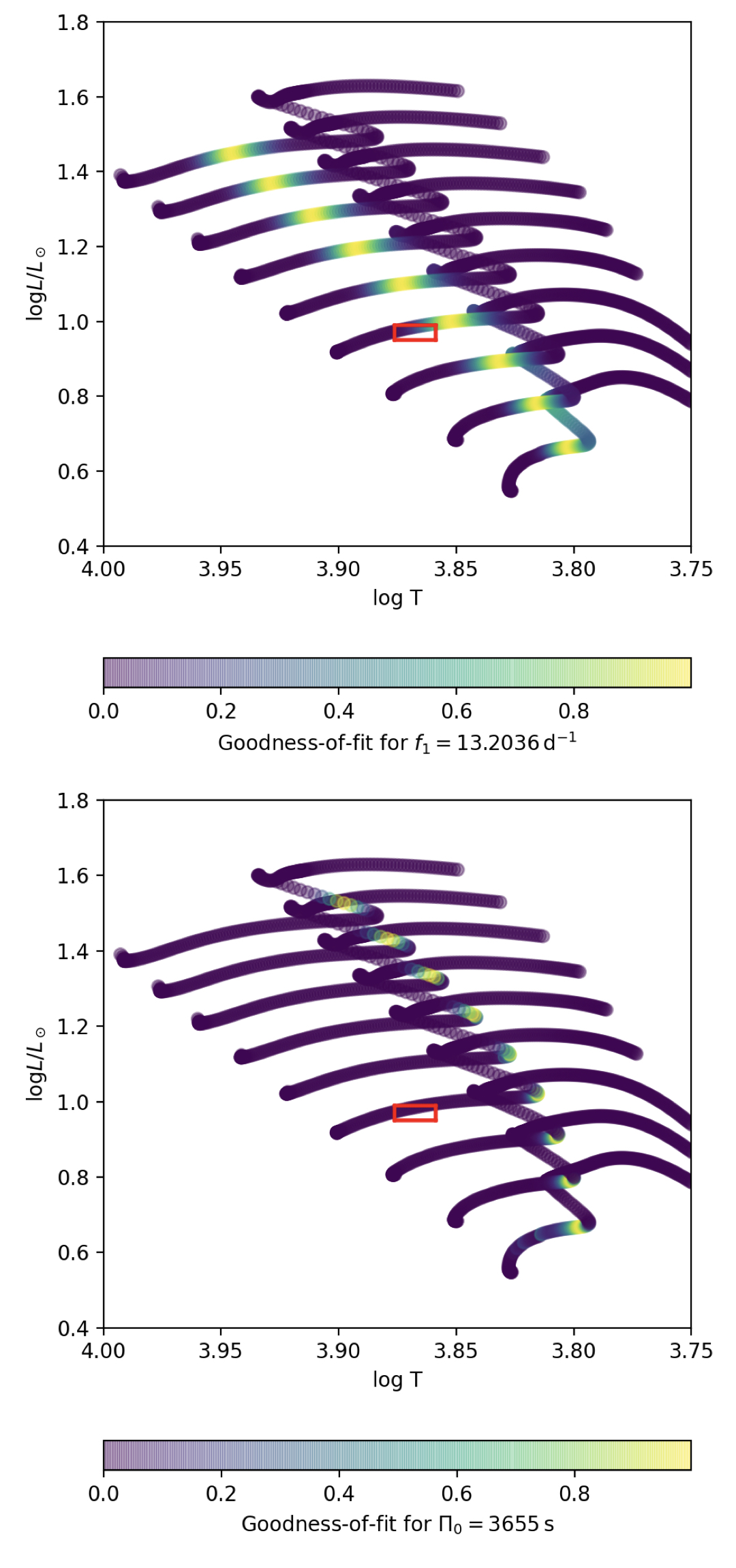}
    \caption{The comparison between the observations and models with $Z=0.020$, mixing length $\alpha=2$, and exponential overshooting $f_\mathrm{ov}=0.02$. The mass is from 1.4 to 2.2 solar masses with steps of 0.1 solar masses. The red box shows the observed region ($T_\mathrm{eff}=7371\pm150\,\mathrm{K}$, $\log L/L_\odot=0.97\pm0.02$), and the color shows the goodness-of-fit for the fundamental mode ($f_1$, top) and asymptotic spacing ($\Pi_0$, bottom), which are proportional to $\exp\left[{-\left(f_\mathrm{1, cal}-f_1\right)^2}\right]$ and $\exp\left[{-\left(\Pi_\mathrm{0, cal}-\Pi_0\right)^2}\right]$, in which $f_\mathrm{1, cal}$ and $\Pi_\mathrm{0, cal}$ are the model-derived fundamental mode and asymptotic spacing.}
    \label{fig:two_HR}
\end{figure}

\section{Binary nature}
By using the well-known `Observed minus Calculated' ($O-C$) method \citep{Sterken2005}, we measured the light maximum times of the fundamental mode and found a significant time shift, as shown in Fig.~\ref{fig:O-C}. The light maximum times vary from $-$200 seconds to about 100 seconds, and the maxima of the $O-C$ diagram appear at about 1800 and 12000 cycles, leading to a period of around 800 days. We deduce that the periodic variation of the light maximum time is caused by the orbital motion of the pulsating star TIC\,308396022, as observed in many previous large-amplitude variable stars \citep[e.g.][]{Fu_2003, Conroy2014, Li_2018}. 

To investigate the orbital parameters, we applied the phase-modulation method to the fundamental mode $f_1$ and its harmonics 2$f_1$, 3$f_1$, and 4$f_1$ \citep{2014MNRAS.441.2515M,Murphy2015,Murphy2016,Hey2020_1, Hey2020_2}, whose phase changes indicated the light travel effect is due to the orbital motion hence allow us to determine the orbital parameters. Figure~\ref{fig:time_delay} displays the time delay and the best-fitting model. The medians and one-sigma uncertainties from the positions of a Monte Carlo Markov Chain (MCMC) give the orbital period $803.5\pm0.6\,\mathrm{d}$, the projected semi-axis $a_1\sin i /c = 262.9\pm1.2 \,\mathrm{s}$, the eccentricity $e=0.015\pm0.006$, and the mass function $f\left(M_1, M_2, \sin i\right)=\frac{\left(m_2\sin i\right)^3}{\left(m_1+m_2\right)^2}=0.0302\pm0.0004\,\mathrm{M_\odot}$. Assuming the mass of TIC\,308396022 is $1.5\,\mathrm{M_\odot}$ ($2.5\,\mathrm{M_\odot}$), the mass of the secondary is $0.49\,\mathrm{M_\odot}$ ($0.67\,\mathrm{M_\odot}$) for the inclination $i=90^\circ$, or $0.58\,\mathrm{M_\odot}$ ($0.8\,\mathrm{M_\odot}$) for $i=60^\circ$, which is within the typical mass range for a C/O WD \citep{2010A&ARv..18..471A,2019A&ARv..27....7C}.

To investigate the possible evolutionary history of this system, an evolutionary population synthesis simulation was conducted, by using the code developed initially by \citet{hurley00,hurley02} and recently updated by \citet{zuo14c}. To simulate this system ($M_1\sim1.5-2.5\,\mathrm{M_\odot}$; $M_2\sim0.49-0.8\,\mathrm{M_\odot}$; $P_{\rm orb}\sim 800$ days), the progenitor was found to have these features: mass of primary in $\sim 1.7-2.8\,\mathrm{M_\odot}$, secondary $\sim 1.4-2.2\,\mathrm{M_\odot}$, and the initial orbits $\sim 200-600\,\mathrm{R_\odot}$. Taking a typical system for example, we chose a primordial binary system with masses $2.3\,\mathrm{M_\odot}$ (the primary) and $1.81\,\mathrm{M_\odot}$ (the secondary) in a $413\,\mathrm{R_\odot}$ orbit. 
The primary evolves and first arrives at its asymptotic giant branch (AGB) stage (at 1011 Myr), the radius of the expanding AGB star exceeds its Roche lobe (RL) and it starts mass-transfer to its companion \citep{1970A&A.....7..150L,2000MNRAS.316..689K,2014MNRAS.441..354T}. 
The orbit circularizes during the mass transfer process and then the system leaves a post-AGB star, and in this case, a $0.57\,\mathrm{M_\odot}$ C/O WD with 
a $1.96\,\mathrm{M_\odot}$ (rejuvenated) MS star in a $504\,\mathrm{R_\odot}$ orbit, which is in line with the orbital parameter of this system. 
We note WDs are often found in low-eccentricity systems, and for A/F stars that are the primaries of systems in the period range detectable by \textit{Kepler} (100-1500 d), 21\% of the companions are WDs \citep{Murphy2018}.
After that, the rejuvenated secondary evolves to expand and fills its RL on the AGB (at 1709 Myr). 
Then the binary may enter into a common envelope \citep[see][for reviews]{Ivanova13}, leaving a double WD if it survives.

We have calculated a coarse grid to construct a preliminary model of the pulsator under a single-star evolution, using the stellar evolution code MESA \citep[v12778,][]{Paxton2011ApJS, Paxton2013ApJS, Paxton2015ApJS, Paxton2018ApJS, Paxton2019ApJS} and the stellar oscillation code GYRE \citep[e.g.][]{Townsend2013MNRAS}. The parameters of the grid were: mass from 1.4 to 2.2 solar masses with step of 0.1 solar masses, metallicity $Z=0.020$, mixing length $\alpha = 2$ and exponential overshooting $f_\mathrm{ov}=0.02$. The results are shown in Fig.~\ref{fig:two_HR}, where the color represents the goodness-of-fit for the fundamental mode ($f_1$, top panel) and the g-mode asymptotic spacing ($\Pi_0$, bottom panel). We find that within the observation region ($T_\mathrm{eff}=7371\pm150\,\mathrm{K}$, $\log L/L_\odot=0.97\pm0.02$ derived in Section~\ref{sec:locationHR}), there is no model that satisfies both the p- and g-mode pulsations. We also tried several other selections of metallicity ($Z=0.0043,~ 0.0054, ~0.0068, ~0.0085, ~0.010, ~0.013, ~0.016, ~0.025$), but still could not find a compatible model. The absence of a compatible model implies that the star has different core and envelope formation history, perhaps due to mass transfer from the secondary star. Some recent researches show that asteroseismology has the ability to detect stellar merger or mass transfer in red giants \citep{Rui2021, Deheuvels2021}, and we suggest that TIC\,308396022 might be a MS counterpart showing mass transfer by asteroseismology. We also note that \citet{2021MNRAS.505.3206M} perform binary-evolution and pulsational modellings of KIC 10661783, an $\delta$\,Sct--$\gamma$\,Dor hybrid in an eclipsing binary system, and show that a mass transfer has a significant impact on the g-mode excitations.  

To confirm the history of the mass transfer of TIC 308396022, we propose two plans for future work: 
\begin{itemize}
    \item Performing detailed seismic modelling including the mass transfer from companions, since mass transfer may lead to incompatibility between the g- and p-mode pulsations and we cannot find a normal single-star evolution model that fits both the g and p modes in the same time; 
    \item Conducting spectroscopic observations, which may help to indicate an enhancement of s-process elements which are created during the AGB phase of the donor's evolution.
\end{itemize}

\section{Conclusions}

We analyzed the pulsating behavior of TIC\,308396022 by using the three-year photometric observations delivered from the TESS mission, and extracted p- and g-mode pulsations from the 2-min cadence data with Fourier transform. The strongest peak $f_1=\ponestring\,\mathrm{d^{-1}}$ is identified as the radial fundamental mode, and we also detect the first overtone at $f_2=\ptwostring\,\mathrm{d^{-1}}$ and the second overtone $f_3=\pthreestring\,\mathrm{d^{-1}}$, which leads this star to be a new radial triple-mode HADS. Another strong mode at $f_4$ = 18.96664(1) $\mathrm{d^{-1}}$ is probably an $l$ = 1 mode.

In the low-frequency region, we find an equally-spaced period pattern with the period spacing $\Delta P = 2460\,\mathrm{s}$. The asymptotic spacing is measured as $\Pi_0=3655\pm13\,\mathrm{s}$, lower than most of the $\gamma$\,Dor stars in \citet{LiGang2019}. We propose that TIC\,308396022 is the first radial triple-mode HADS with a g-mode period spacing pattern. 

TIC\,308396022 also shows clear orbital motion since its light maxima times and pulsation phases vary periodically. The orbit, which has a long period ($803.5\pm0.6\,\mathrm{d}$), a very small eccentricity (0.015), and a small mass function, indicates that the system probably has undergone mass transfer and the companion is likely to be a C/O WD. Based on the derived luminosity log $L/L_{\odot}$ = 0.97 $\pm$ 0.02, and the effective temperature $T_\mathrm{eff}$ = 7371 $\pm$ 150\,K, TIC\,308396022 lies near the bottom of the $\delta$\,Sct instability strip. However, we were not able to find a theoretical model that matches the observed properties (L, T$_{eff}$, $f_1$and $\Pi_0$). Spectroscopic observations and detailed evolutionary--seismic modellings considering the mass accretion are needed to reveal the history of this system further.

\section*{Acknowledgments}

This research is supported by the program of the National Natural Science Foundation of China (grant Nos. 11573021, U1938104 and 12003020). Gang Li acknowledges support from the project BEAMING ANR-18-CE31-0001 of the French National Research Agency (ANR) and from the Centre National d'Etudes Spatiales (CNES). We are grateful to the Australian Research Council for support (DP 210103119). This work has made use of data from the European Space Agency (ESA) mission
{\it Gaia} (\url{https://www.cosmos.esa.int/gaia}), processed by the {\it Gaia}
Data Processing and Analysis Consortium (DPAC,
\url{https://www.cosmos.esa.int/web/gaia/dpac/consortium}). Funding for the DPAC
has been provided by national institutions, in particular the institutions
participating in the {\it Gaia} Multilateral Agreement. We would like to thank the TESS science team for providing such excellent data.

\end{document}